# Fast Interlayer Energy Transfer from the *Lower Bandgap* $MoS_2$ to the *Higher Bandgap* $WS_2$

**Gayatri,[1] Mehdi Arfaoui,[2] Debashish Das,[3] Tomasz Kazimierczuk,[1] Sabrine Ayari,[4] Natalia Zawadzka,[1] Takashi Taniguchi,[5] Kenji Watanabe,[6] Adam Babiński,[1] Saroj K. Nayak,[3] Maciej R. Molas,[1] Arka Karmakar[1*]**

[1] University of Warsaw, Faculty of Physics, 02-093 Warsaw, Poland

[2] Université Paris-Saclay, ONERA, CNRS, Laboratoire d'étude des microstructures (LEM), F-92322 Châtillon Cedex, France

[3] School of Basic Sciences, Indian Institute of Technology Bhubaneswar, Khordha 752050, Odisha, India

[4] De Vinci Higher Education, De Vinci Research Center, Paris, France

[5] Research Center for Materials Nanoarchitectonics, National Institute for Materials Science, Tsukuba, Ibaraki 305-0044, Japan

[6] Research Center for Electronic and Optical Materials, National Institute for Materials Science, Tsukuba, Ibaraki 305-0044, Japan

* arka.karmakar@fuw.edu.pl; karmakararka@gmail.com

**Abstract:**

Energy transfer (ET) is a dipole-dipole interaction, mediated by the virtual photon. Traditionally, ET happens from the higher (donor) to lower bandgap (acceptor) material. However, in some rare instances, ET can happen from the lower-to-higher bandgap material, depending on the strong overlap between the acceptor photoluminescence (PL) and the donor absorption spectra. In this work, we report an ET process from the lower bandgap $MoS_2$ to the higher bandgap $WS_2$, due to a near 'resonant' overlap between the $MoS_2$ B and $WS_2$ A excitonic levels. Changing the $MoS_2$ bandgap from direct-to-indirect by increasing the layer number results in a reduced ET rate, evidenced by the quenching of the $WS_2$ PL emission. Our work shows at 300 K, the ET timescale of ~33 fs is faster than the reported thermalization of the $MoS_2$ excitonic intervalley scattering (K↔K') time and competing with the ultrafast charge transfer timescale.

Thus, allowing us to open a new direction in understanding the competing inter/intralayer processes.

**Introduction:**

Van der Waals (vdW) heterostructures (HSs) made of the layered transition metal dichalcogenide (TMDC) materials have shown huge potential in designing next-generation optoelectronic device applications.[1–4] Placing materials into the atomically closed proximity in the TMDC HSs are well established by previous studies.[1,5] Materials in closed proximity interact *via* near-field interactions, resulting in an energy transfer (ET) (alternatively known as Förster Resonance Energy Transfer, FRET) from one material donor (Dn) to the other acceptor (Ac).[6] The ET process is the backbone of many chemical and biological applications.[7,8] Most importantly, photosynthesis in the plants is based on it.[9] Furthermore, several groups have already reported the presence of the ET process in the newly emerging low-dimensional materials, including the TMDC HSs.[10–17] Thus, considering as the building block in the post-silicon era optoelectronic applications, understanding the ET process in the TMDC HSs at the microscopic level is extremely important for designing better device performances.[18]

The ET process is a dipole-dipole interaction, mediated by the virtual photon.[19] Thus, ET is a long-range interaction compared to the other interlayer processes (charge and Dexter transfer).[18] In these layered HSs, the ET process can survive up to tens of nm.[13,20] Previous works,[21,22] predicted that the interlayer ET coupling in the vdW HSs has a strong dependence on the in-plane center of mass momentum $\hbar Q$ of the photoexcited excitons (*e-h* pairs) and vanishes at $Q \approx 0$. This implies that the carrier lifetime within the light cone is not dependent on the ET rate and ET is dominated by the excitons at larger momenta, *i.e.*, outside of the light cone.

Experimental studies in the field of vdW HSs are mainly focused on a traditional ET process from the higher-to-lower bandgap materials (*i.e.*, thermodynamically favorable),[10–12,14,16] with a reported timescale between the hundreds of fs to few ps, or more.[12,14,16] Usually, the higher bandgap material is the Dn, and the lower bandgap is the Ac. However, in some atypical cases, ET can happen from the Ac-to-Dn material, depending on the strong overlap between the Ac photoluminescence (PL) and the Dn absorption spectra.[23,24] Due to the strong overlap between the bands and associated difficulties in spectrally resolving the emissions, experimental examinations on this type ET process remain in scarce.[25] The problem associated with finding the 'true' ET timescale continue to exist, due to the limited instrumental capabilities to find the ultrafast inter-/intravalley scattering time at different excitations.

The focus of this work is to find the ET rate/timescale from the Ac-to-Dn material and the effect of bandgap tuning on it. We report an ET process from the lower bandgap 1L molybdenum disulfide ($MoS_2$) to the higher bandgap 1L tungsten disulfide ($WS_2$) separated by a thin charge-blocking hexagonal boron nitride (hBN) interlayer, due to a near 'resonant' overlap between the $MoS_2$ B and $WS_2$ A excitonic levels. We also want to mention that we do not consider an ET from the $WS_2$ A to $MoS_2$ A excitonic level for a discussion. As mentioned earlier, this type of higher-to-lower bandgap ET was extensively studied in previous reports, including this same type-II HS configuration.[12,16] We then expand this work by investigating the effect of $MoS_2$ direct-to-indirect bandgap crossover on the ET process by changing the $MoS_2$ layer thickness,[26,27] while keeping the interlayer separation constant. We study this ET process by measuring the $WS_2$ PL enhancement/quenching in the HS area as compared to the isolated region. Finally, we find a fast ET timescale of ~33 fs, dominating the ultrafast $MoS_2$ excitonic intervalley scattering time (K↔K'). We employ different room (RT, $T$ = 300 K) and low temperature (LT, $T$ = 5 K) optical spectroscopic techniques; such as differential reflection contrast (RC), PL, photoluminescence excitation (PLE) and time resolved PL (TR-PL), complemented by the advanced theory calculations to justify the experimental observations.

**Results:**

**Absorption measurement**

RT absorption spectra converted from the RC measurements (see Experimental Details) show a near 'resonant' overlap between the $MoS_2$ B and $WS_2$ A excitonic levels (shaded region in Figure 1a). This agrees well with the earlier report.[28] The main concept of this work is depicted in Figure 1b: In $MoS_2$, 1L-to-multilayer transition opens up an indirect bandgap between the K-Γ valleys.[26,27] Resonant excitation matching with the $MoS_2$ B level creates *e-h* pairs in the K valley. At RT, these 'hot' carriers (*e-h* pairs) can instantaneously (tens of fs) scatter to the Λ or Γ valley, respectively.[29] The carriers remain in the $MoS_2$ B excitonic level can 'resonantly' couple to the $WS_2$ A level *via* the ET process[25] and give rise to the $WS_2$ PL emission in the HS area. The ET efficiency will depend on the available *e-h* pairs population at the $MoS_2$ K valley. Thus, it is expected to see the maximum $WS_2$ PL enhancement for HS with 1L $MoS_2$ and a reduction in $WS_2$ PL intensity with the increasing $MoS_2$ layer number.

**Photoluminescence excitation spectroscopy**

To experimentally verify the above mentioned concept, we prepare a HS sample with 1L $WS_2$ on top and nL $MoS_2$ (n = 1, 2, 4 & 5) as the bottom layer with a thin hBN interlayer with a thickness of about ~5.5 nm – to completely suppress the interlayer charge transfer process[30] (see supplementary Figure S1 and Experimental section for the details). Figure 2a represents the graphical illustration of this sample. The RT PLE false colormap of the $WS_2$ excitonic emission shows a significant increase in the 1L HS region and then a gradual decrease with the increasing $MoS_2$ layer number, as compared to the isolated $WS_2$ region (Figure 2b). Note that the naming of the HS regions (1L, 2L, etc.) is based on the corresponding $MoS_2$ layer thickness and all the PLE maps in Figure 2b are presented with the same intensity range for a better visual comparison. A cut at the 2.4 eV excitation reveals ~3× enhancement of the $WS_2$ PL emission from the 1L HS area compared to the isolated $WS_2$ region (Figure 2c). This enhancement then drops to only ~1.25× for the 2L HS and < 1 for the 4L and 5L HSs. A vertical cut at the $WS_2$ emission energy ~2 eV (along the dotted line in Figure 2b) provides the PLE spectrum, *i.e.*, the emission intensity profile as a function of the excitation energy. Dividing the HS $WS_2$ emission by the isolated $WS_2$ emission gives the PL enhancement factor as a function of the excitation energy (Figure 2d). Enhancement factor > 1 and < 1 meaning enhancement and quenching in the HS $WS_2$ emission, respectively. The 1L HS region shows an overall increasing enhancement with the excitation energy. However, the 2L HS emission shows a quenching at the beginning and then after ~2.17 eV excitation it shows an enhancement with the increasing excitation energy. The 4L and 5L regions reveal a quenched emission throughout the entire excitation. To verify the data reproducibility, we prepare a new sample and find a similar ~2× pronounced 1L HS $WS_2$ emission (Figure S2) at an excitation energy of 2.1 eV. The LT PL spectra taken under 2.4 eV excitation show a slight $WS_2$ neutral exciton ($X_0$) enhancement (~1.3×) from the 1L HS area and all other HS regions show a quenched emission (Figure 2e). A detailed discussion of these experimental results is presented in the following sections.

Strategically in the PLE experiment, we used an excitation range from close to the $WS_2$ A exciton to 2.4 eV (close to B excition) only (Figure 2b). So, the *e-h* pair excitation and generation can be restricted only within the K valley and the 'band-nesting' regions (*i.e.*, high-lying excitonic states) do not participate in the ET process. This is due to the fact of avoiding any contribution from the fast intervalley electron scattering ($\Lambda \rightarrow K$),[25] exceeding our experimental temporal resolution. The overall enhancement of the 1L HS PL emission suggests an ET process.[25] This is further supported by an increase in the 1L HS PL linewidth (Figure S3 and Table S1) compared to the $WS_2$.[11] As the ET rate approaches to zero for $Q = 0$, it is not expected that the ET coupling influences the excitons within the light cone.[21,22] Thus, considering the relatively large interlayer distance (~5.5 nm hBN thickness), we conclude a long-range dipolar coupling is responsible for the HS PL enhancement and we ignore the other short-range interlayer processes, such as charge and Dexter transfer.[11,30]

The indirect bandgap between the K-Γ valleys in nL $MoS_2$ decrease with the increasing layer number[26,27] (see Figure S4 description for the computational details). However, with the increasing $MoS_2$ layer number, the following other two things also happen: the gap between the top of the Γ-K valleys in the valence band (VB) increases[26] (Figure S4) and the hole effective mass also significantly increases.[31] The heavier hole effective mass directly reduces the carrier mobility.[32] Thus, the hole transfer back to the K valley becomes less probable with the increasing $MoS_2$ layer number. Due to this, a lesser number of *e-h* pairs in the $MoS_2$ K valley can participate in the ET process, and the PL enhancement factor (*i.e.*, ET efficiency) starts to drop with the increasing $MoS_2$ layer number (Figure 2d). This also explains the behavior observed in the LT data (Figure 2e). The ET rate decreases with the decreasing temperature,[15] and at cryogenic temperature carriers scattering back to the K valley from the Λ or Γ valley become less probable due to the reduced electron-phonon interaction.[29] Thus, at 5 K, despite having similar overlaps between the $WS_2$ A and $MoS_2$ B levels (Figure S5), we see a minor enhancement of the $WS_2$ $X_0$ emission in the 1L HS and quenching for other HS regions. Whereas, at RT, due to the available large phonon numbers, the probability of these 'hot' carriers returning back to the K valley increases.[33,34] As a result, $WS_2$ PL increases at RT. Later, we consider different phonon-mediated valley scattering processes in our analysis.

**Transition dipole moment calculation**

Now, we take into account the transition dipole moment (TDM) strength of the direct K-K $MoS_2$ and $WS_2$ A and B excitonic states. Figure 3 illustrates the calculated squared TDM at the K valley using the density functional theory (DFT) at the independent particle level (details in Figure S6). The A and B transitions are represented correspondingly with the red and black bars. The dipole transition strength in $WS_2$ is significantly stronger than in nL $MoS_2$. Since the dipole transition strength is inversely proportional to this energy difference, therefore, we have a large difference in A and B transitions strength for $WS_2$ (Figure S7). Whereas for the case of $MoS_2$, the strength of A and B transitions is relatively close. The number of A and B transitions increases with the number of layers in the material.[35] A pair of both A and B transitions exist in materials like 2L $MoS_2$, but both the A-type transitions occur at nearly identical energies, making them difficult to distinguish. Similarly, in 4L $MoS_2$, four A-type transitions show a similar degeneracy. A higher TDM value corresponds to a higher oscillator strength.[36] Thus, the A-type transitions are inherently stronger than the B-type transitions. This difference, combined with the slightly higher $MoS_2$ B energy position, results in an ET process from the $MoS_2$ B to $WS_2$ A excitonic level.

**Time resolved photoluminescence experiment**

In order to get more insight into the dynamics, we study TR-PL emission of the sample. TR-PL spectra taken at 300 K show a similar faster decay ~10-11 ps (details in the Experimental section) for both the $MoS_2$ and $WS_2$ excitonic emission in the 1L HS, and $WS_2$ emission in the 2L HS (Figure 4a). The faster time constant provides the intrinsic PL lifetime and the slower component is associated with the trap states transition.[10,13] A similar excitonic decay time in the 1L HS for both materials suggests that this reported ET does not result from a prolonged Dn lifetime.[14] Interestingly, we observe a decrease in the $WS_2$ lifetime to ~4 ps from the isolated region (Figure S8). A more than 2.5× increase in the $WS_2$ lifetime at the HS region suggests a possible 'both-way' excitonic transfer between the $MoS_2$ B and $WS_2$ A level (outside of the radiative light cone[37]) *via* ET process, before a radiative recombination happens at the $WS_2$ K valley. Due to a similar resonant overlap between two excitonic states, earlier we observed an identical prolonged HS PL lifetime in a twisted $MoSe_2$ homobilayer study.[34] Finding the mechanism behind the different $WS_2$ rise time in the 1L and 2L HSs (Figure 4a) is beyond the scope of this present work.

**Theoretical modeling of energy transfer**

For investigating the ET process within this 1L HS, we theoretically calculate the time- and energy-resolved PL spectra using the Elliot formula,[38] which is directly proportional to the bright exciton population $N_\eta^{B,\lambda}$ within the light cone:

$$I_{TR-PL}(\omega, t) \; = \; \sum_{\eta \in \{Dn, Ac\}} \sum_\lambda N_\eta^{B,\lambda}(t) \; \left|f_\lambda^{X,\eta}\right|^2 \mathcal{L}_\eta^\lambda(\omega) \qquad (1)$$

with the Lorentzian line-shape $\mathcal{L}_\eta^\lambda(\omega) = \left( \left[\hbar\omega - E_\eta^\lambda(0)\right]^2 + {\Gamma_{tot}^{\lambda,\eta}}^2 \right)^{-1}$, $N_\eta^{B,\lambda}(t)$ is the time-dependent excitonic population after excitation, $\left|f_\lambda^{X,\eta}\right|^2 \propto \frac{2\, m_0\, E_\eta^\lambda(0)}{\hbar^2} \left|\sum_{\mathbf{k}_{\eta,rel}} \varphi_{\eta,\lambda}^{rel}\left(\mathbf{k}_{\eta,rel}\right)\right|^2$ is the oscillator strength

of the $\eta \in \{Dn, Ac\}$ exciton with state $\lambda$, and $E_{\eta}^{\lambda}(0)$ is the energy and relative-motion wavefunction $\varphi_{\eta,\lambda}^{rel}\left(\mathbf{k}_{\eta,rel}\right)$ with momentum $\hbar \boldsymbol{k}_{\eta,rel}$ (see Figure S9 for details). $\Gamma_{tot}^{\lambda,\eta} = \Gamma_{rad}^{\lambda,\eta} + \Gamma_{inh}^{\lambda,\eta}$ denotes the broadening describing the radiative and non-radiative decay rate and it was taken from our experiment. A false colormap of the time- and energy-dependent PL emission from the 1L HS at 0 K is shown in Figure 4b, where the inset presents the evolution of the 1Ls $MoS_2$ B and $WS_2$ A states' emissions at an early timescale. This calculation is based on the assumption that at $t = 0$, the exciton population is present only in the Dn $MoS_2$ B state. With the increasing time, the ET process starts to contribute to an increase in the Ac $WS_2$ A state emission, resulting in a vanishing $MoS_2$ emission (Figure 4c). Further increasing the time results in decreased $WS_2$ emission. This behavior is attributed to the radiative and nonradiative recombination of excitons in $WS_2$. Which results the PL intensity being progressively vanished beyond 4 ps (Figure 4b). Recently, a similar behavior was reported in an organic-TMDC HS.[15]

Our model yields the complete excitonic series, but in this work we restrict our discussion to the fundamental 1s state, since the 2s excited level lies about ~150 meV above it and thus contributes negligibly to the ET dynamics. So, we focus only on the energetically low-lying exciton states where the ET is more efficient, specifically for the direct K/K' 1s-excitons (*i.e.,* near-resonance condition and higher oscillator strength lead to stronger spectral overlap) for 1L HS (since for $n > 1$, the ET become less efficient) and we do not take into account an ET from the $WS_2$ to $MoS_2$ layer (Ac→Dn). Based on the Fermi's Golden rule,[38] the exciton-mediated ET rate $\langle\Gamma_{\mathrm{F}}\rangle^{Dn\to Ac}(z,T)$ at finite temperature T can be calculated as the thermal average from all Dn excitonic states to Ac as follows

$$\langle\Gamma_F\rangle^{Dn\to Ac}(z,T) = \langle 1/\tau_{F,\mathrm{eff}}^{Dn\to Ac}\rangle(z,T) = \sum_{\lambda,\lambda'}\sum_{Q} f_{\lambda}^{Dn}(\mathbf{Q},T)\, \zeta_{\lambda\to\lambda'}^{Dn,Ac}(\boldsymbol{Q},z) \qquad (2)$$

$$\zeta_{\lambda\to\lambda'}^{Dn,Ac}(\mathbf{Q},z) = \frac{2\pi}{\hbar}\left|M_{\lambda\to\lambda'}^{Dn,Ac}(\mathbf{Q},z)\right|^2 \delta\left(E_{\lambda}^{\mathrm{Dn}}(\mathbf{Q}) - E_{\lambda'}^{\mathrm{Ac}}(\mathbf{Q})\right);$$

Here, in the low-density limit, the thermal occupation factor expressed by the Boltzmann distribution $f_{\lambda}^{Dn}(\mathbf{Q},T) = \exp\left[-E_{\lambda}^{Dn}(\mathbf{Q})/k_B T\right]/Z$ with the partition function $Z = \sum_{\lambda}\sum_{Q}\exp\left[-E_{\lambda}^{Dn}(\mathbf{Q})/k_B T\right]$. Here, $\mathrm{E}_{\lambda(\lambda')}^{\mathrm{Dn(Ac)}}(\mathbf{Q})$ is the Dn (Ac) exciton energy dispersion at state $\lambda(\lambda')$ with center of mass momentum $\hbar\mathbf{Q}$. The delta function ensures energy conservation during the transfer process. The Förster coupling matrix element $M_{\lambda\to\lambda'}^{Dn,Ac}$, represents the dipole-dipole interaction between the Dn and Ac excitons (see Figure S10).

Solving the Equation 2, gives the ET rate ~20 ps$^{-1}$ at 5K. This results in ET timescale τ ~50 fs for the 1L HS. Notably, in the 1L HS, around 50 fs we also see an onset of the $WS_2$ emission overtaking the $MoS_2$ emission (inset of Figure 4b). Which indicates the ET process to take over the carrier relaxation channels. Both of these similar findings, despite using very different approaches, justify the calculated ET timescale. To calculate the ET timescale at RT, we use Equation 2, and the obtained value is consistent with the population dynamics observed in the experimental TR-PL data (see Figure S11), where the calculated and measured trends show excellent agreement. This yields a fast ET timescale of approximately 33 fs at 300 K. A similar trend, *i.e.*, an increasing ET rate with the increasing temperature was also reported earlier.[15,39]

Having calculated the Dn to Ac ET rate, we now incorporate it into the coupled rate equations to describe the TR-PL behavior from the $MoS_2$ $B_{1s}$ $K^{\downarrow}$-$K^{\downarrow}$ to $WS_2$ $A_{1s}$ $K^{\uparrow}$-$K^{\uparrow}$ state, by using the following equations:

$$
\begin{aligned}
\frac{dN_{Dn}^{B,K}(t)}{dt} &= -\Gamma_{Dn}^{B,K} N_{Dn}^{B,K} + \Gamma_{Dn}^{DB,K} N_{Dn}^{D,K} - \Gamma_{Dn}^{BD,K} N_{Dn}^{B,K} - \Gamma_{F}^{Dn(B,K)\to Ac(B,K)} N_{Dn}^{B,K} + \Gamma_{F}^{Ac(B,K)\to Dn(B,K)} N_{Ac}^{B,K} - \\
&\quad \sum_{\chi\in\{\Gamma,\Lambda,K'\}} \Gamma_{ph,Dn}^{K\chi} N_{Dn}^{B,K} + \sum_{\chi\in\{\Gamma,\Lambda,K'\}} \Gamma_{ph,Dn}^{\chi K} N_{Dn}^{D,\chi} + g_{Dn}^{B,K}(t) \\
\frac{dN_{Dn}^{D,K}(t)}{dt} &= -\Gamma_{Dn}^{D,K} N_{Dn}^{D,K} - \Gamma_{Dn}^{DB,K} N_{Dn}^{D,K} + \Gamma_{Dn}^{BD,K} N_{Dn}^{B,K} \\
\frac{dN_{Dn}^{D,\Gamma}(t)}{dt} &= -\Gamma_{Dn}^{D,\Gamma} N_{Dn}^{D,\Gamma} + \Gamma_{ph,Dn}^{K\Gamma} N_{Dn}^{B,K} - \Gamma_{ph,Dn}^{\Gamma K} N_{Dn}^{D,\Gamma} \\
\frac{dN_{Dn}^{D,\Lambda}(t)}{dt} &= -\Gamma_{Dn}^{D,\Lambda} N_{Dn}^{D,\Lambda} + \Gamma_{ph,Dn}^{K\Lambda} N_{Dn}^{B,K} - \Gamma_{ph,Dn}^{\Lambda K} N_{Dn}^{D,\Lambda} \\
\frac{dN_{Dn}^{D,K'}(t)}{dt} &= -\Gamma_{Dn}^{D,K'} N_{Dn}^{D,K'} + \Gamma_{ph,Dn}^{KK'} N_{Dn}^{B,K} - \Gamma_{ph,Dn}^{K'K} N_{Dn}^{D,K'} \\
\frac{dN_{Ac}^{B,K}(t)}{dt} &= -\Gamma_{Ac}^{B,K} N_{Ac}^{B,K} - \Gamma_{Ac}^{BD} N_{Ac}^{B,K} + \Gamma_{Ac}^{DB} N_{Ac}^{D,K} - \Gamma_{F}^{Ac(B,K)\to Dn(B,K)} N_{Ac}^{B,K} + \Gamma_{F}^{Dn(B,K)\to Ac(B,K)} N_{Dn}^{B,K} - \\
&\quad \sum_{\chi\in\{\Lambda,K'\}} \Gamma_{ph,Ac}^{K\chi} N_{Ac}^{B,K} + \sum_{\chi\in\{\Lambda,K'\}} \Gamma_{ph,Ac}^{\chi K} N_{Ac}^{D,\chi} + g_{Ac}^{B,K}(t) \\
\frac{dN_{Ac}^{D,K}(t)}{dt} &= -\Gamma_{Ac}^{D,K} N_{Ac}^{D,K} - \Gamma_{Ac}^{DB,K} N_{Ac}^{D,K} + \Gamma_{Ac}^{BD,K} N_{Ac}^{B,K} \\
\frac{dN_{Ac}^{D,\Lambda}(t)}{dt} &= -\Gamma_{Ac}^{D,\Lambda} N_{Ac}^{D,\Lambda} + \Gamma_{ph,Ac}^{K\Lambda} N_{Ac}^{B,K} - \Gamma_{ph,Ac}^{\Lambda K} N_{Ac}^{D,\Lambda} \\
\frac{dN_{Ac}^{D,K'}(t)}{dt} &= -\Gamma_{Ac}^{D,K'} N_{Ac}^{D,K'} + \Gamma_{ph,Ac}^{KK'} N_{Ac}^{B,K} - \Gamma_{ph,Ac}^{K'K} N_{Ac}^{D,K'} \qquad (3)
\end{aligned}
$$

where, $N_{Dn}^{B(D),\chi}$ and $N_{Ac}^{B(D),\chi}$ denote the $\chi = \{K, K', \Lambda, \Gamma\}$-valley bright (B) and momentum-forbidden dark (D) exciton populations in the $MoS_2$ B and $WS_2$ A states, respectively. $\Gamma_{Dn(Ac)}^{B(D),\chi}$ are their corresponding radiative decay rate. $\Gamma_{Dn(Ac)}^{DB,K}/\Gamma_{Dn(Ac)}^{BD,K}$ are the intravalley spin forbidden bright-dark scattering rates (with $\Gamma_{Dn(Ac)}^{BD(DB),K} = \Gamma_{Dn(Ac)}^{DB(BD),K} e^{-|\Delta E|/k_B T}$; $\Delta E = E_{1s}^{B,K}(0) - E_{1s}^{D,K}(0)$, see details in the Supplementary Information). $\Gamma_{ph,\eta}^{K\chi'} = \gamma_{ph,\eta}^{K\chi'}(1 + n_{\mathbf{q}\vartheta})$ correspond to the emission intervalley phonon-assisted scattering rate between the K-valley bright state and the momentum indirect $\chi'$ valley dark reservoir $N_{\eta}^{D,\chi'}$ in the $\eta$ layer. $\gamma_{ph,\eta}^{\mathrm{K}\chi'}$ is the zero temperature relaxation rate and $n_{\mathbf{q}\vartheta} = \left(e^{\Delta \mathrm{E}_{\mathrm{K}\chi'}/k_B T} - 1\right)^{-1}$ is the Bose Einstein phonon occupation in mode $\vartheta$ and momentum $\hbar\mathbf{q}$. $\Gamma_{ph,Dn}^{\mathrm{K}\Lambda}$ and $\Gamma_{ph,Dn}^{\mathrm{K}\Gamma}$ terms are neglected for 1L $MoS_2$ because the K-Λ and K-Γ excitons sits above the direct K-K exciton, making this transfer thermally insufficient, same for $\Gamma_{ph,Ac}^{\mathrm{K}\Gamma}$. $\Gamma_F^{Dn\rightarrow Ac}$ and $\Gamma_F^{Ac\rightarrow Dn}$ are the Förster interlayer transfer rates. $g_{Dn}^{B,K}(t)$ is the time-dependent Gaussian optical pump profile of the donor state.

Our coupled rate equations model takes nine levels into consideration: the direct A exciton in $WS_2$ and the B exciton in $MoS_2$ (both at the K valley), intravalley spin-forbidden dark states at the K valley for both materials, and momentum-forbidden dark excitons located at the Γ, K′, and Λ valleys. Excitons are assumed to be generated resonantly *via* optical excitation of $MoS_2$ at the K valley. Notably, due to the high exciton binding energies (∼400 meV), we assume that excitons form instantaneously upon excitation, and that exciton ionization into unbound *e-h* pairs is negligible. This assumption can be justified by using the Saha equation,[40] which predicts a very low ionization fraction under these conditions.

By numerically solving the coupled rate equations (Equation 3), we present in Figures S12-S14, the exciton population dynamics at K, K′, Γ, and Λ valleys for both Dn and Ac materials. The corresponding rates values are $\Gamma_{Dn}^{B,K} = 7.2\ ps^{-1}$, $\Gamma_{Ac}^{B,K} = 5.3\ ps^{-1}$, and $\Gamma_{Dn(Ac)}^{D,K} = 10^{-3}\ ps^{-1}$ [41] (see Figures S15-19 and Table S3) The momentum-forbidden dark phonon assisted scattering rates $\Gamma_{ph,\eta}^{\mathrm{K}\chi'}$ are taken from refs.[29,42] Upon resonant optical excitation, the Dn B excitonic state in 1L $MoS_2$ rapidly populates within sub-fs time, then it decays in first few ps which leads to the quenching of the light emission in 1L $MoS_2$ (Figure S12). Indeed, we attribute the short decay of B exciton state in 1L $MoS_2$ to the effective ET to Ac 1L $WS_2$. In fact, in addition to the radiative recombination and ET, these excitons undergo phonon scattering transitions to dark states, either within the intra/inter-valleys. In our study, we specifically focus on the K-Λ, K-Γ and K-K' scattering processes (Figures S12-S13). However, these states lie slightly above the direct bright K-K exciton level typically by tens of meV,[43] rendering thermal population inefficient at LT. Which make the Dn to Ac ET is the most effective channel due to the strong overlap between the acceptor PL and the donor absorption spectra (Figures S13-S14 and S18). Then, the carriers in bright Ac state quickly scatter into the dark Ac states in K' and Λ valleys *via* phonon assisted scattering process, leading to significantly increase of the $N_{Ac}^{\mathrm{D},\Lambda}$, and $N_{Ac}^{\mathrm{D},\mathrm{K}'}$ population and a depletion of the initially Ac bright state population (Figure S17). These dark states lie energetically below the $WS_2$ K-K bright exciton and thus act as effective relaxation channels. The higher final population of Ac excitons in dark states reflects the role of momentum-forbidden transitions being activated through phonon scattering. By suppressing the ET from Dn to Ac channel (Figures S13-S14), the Dn population enhances and decays in a longer time scale, while the population in the Ac decreases significantly, which leads to the quenching of the momentum dark Ac exciton in K' and Λ valley. These results provide compelling evidence that ET, in this scenario, is the primary driver of the observed exciton population transfer and supports the interpretation that ET occurs not only *via* bright-bright channels, but also involves dark-state and multiple valley pathways, in agreement with the full exciton dynamics computed in the coupled rate equations. A details analysis of the role of different scattering channels are presented in Figures S15-19 and S13.

**Discussion:**

In 1L $MoS_2$, the intervalley electron scattering (K↔K') time was reported ~110 fs at resonant A excitation and this value increases to a few ps for excitation matching with the B level.[44] However, it was found that the total valley polarization does not fully relax for ~10 ps due to the slower hole scattering rate.[45] These previous findings complement our observation. It strongly suggests that the ET timescale from the $MoS_2$ B to $WS_2$ A level has to be faster than the $MoS_2$ intervalley scattering (K↔K') time at resonant B excitation. Otherwise, carriers generated at the $MoS_2$ B excitonic level would transfer to the A level in the K valley and that lower energy $MoS_2$ A exciton could not create excess carriers in the higher energy $WS_2$ A level. It is worth mentioning here that the calculated RT ET timescale (~33 fs) in our study is the fastest reported ET rate in any low-dimensional system and comparable with the fast interlayer charge transfer time.[12]

In this paragraph, we take into account of other possible mechanisms to contribute in the HS PL enhancement process. Considering the entire 1L $WS_2$ is placed onto the same hBN flakes (both, interlayer and bottom), we can exclude the possibility of an ET from hBN defect states[46] as a primary source of PL enhancement. Interference in multilayered thin film nanostructure enhances the absorption and PL emission, compared to a pristine layer.[47] However, if this was the case, 4L/5L HSs should emit more than 1L/2L HSs. Thus, we can rule out the possibility of interference in observed HS PL enhancement. Also, comparing the several orders lower excitation power used in our measurements, we can eliminate the possibility of the Meitner-Auger type transfer.[48]

In conclusion, we show that an ET process occurs from the *lower-to-higher* bandgap material due to a near 'resonant' overlap between the 1Ls of $MoS_2$ B and $WS_2$ A excitonic level. With the increasing $MoS_2$ layer thickness, an indirect bandgap opens between the K-Γ valleys. This allows for an instantaneous transfer of the photocarriers to the Λ or Γ valley, resulting in a weaker ET process at the K valley. We show that the ET timescale in 1L HS with the above bandgap excitation is faster than the stabilization of the $MoS_2$ intervalley electron scattering (K↔K'). This work provides a new perspective on understanding the unconventional ET process for developing TMDC HS-based real optoelectronic device applications.

**Methods:**

**HS fabrication**

We obtain the hBN crystals from the National Institute for Materials Science, and commercial $MoS_2$ and $WS_2$ crystals from the HQ Graphene and 2D Semiconductors, respectively. TMDCs and hBN were exfoliated from the bulk crystals using the PDMS-based (Gel-Pak) mechanical exfoliation method.[25] We directly cleave the bottom layer of thick hBN on the $SiO_2$/Si or quartz substrate placed on a 60°C hotplate for 20 min., and then we step-by-step stack the remaining layers using a custom home-built semiautomatic transfer stage. After every step of layering, we perform an annealing process inside an e-beam evaporator chamber under high vacuum (~$7\times10^{-4}$ Torr) at 180°C with Ar flow rate of 100 sccm. for 45 min. to improve the adhesion between the layers and remove any polymer residues from the surface. After the top layer stacking, the sample was annealed under the same condition for 3 hrs.

**Characterization**

The differential RC measurements were performed using a tungsten–halogen lamp focused by a Mitutoyo Plan Apo SL 50× (N.A. 0.42) objective. The sample was mounted on a piezo stage with free movements in x-y-z directions. The signal was directed from the sample to the Princeton spectrometer equipped with 150 grooves/mm, and then detected using a CCD camera. The differential RC is defined as $(R_s-R_{sub})/(R_s+R_{sub})$, where $R_s$ is the reflection intensity from the sample and $R_{sub}$ is the reflection intensity from the substrate. Absorbance spectra were calculated from the RC data as reported in ref.[34,49]

Ambient condition μ-PL/PLE experiments were performed using a broadband supercontinuum laser source connected to a monochromator to control the excitation wavelength. The laser beam was focused on the sample using a Mitutoyo Plan Apo SL 50× (N.A. 0.55) objective with spot of about ~1 μm diameter.

The average power was kept constant for all the wavelengths only at ~3.6 μW using a noise eater to avoid heating up the sample and any high power induced nonlinear effect. For LT measurement, the sample was loaded into a continuous flow cryostat and cooled down with liquid helium (LHe) flow. The gaps in the PL spectra (Figure 2c) are due to the removal of the transmission line (leakage) from a longpass filter at the detection path (Figure S20).

We performed the TR-PL measurements using a S20 synchroscan streak camera. The sample was excited with 450 nm (~2.75 eV) excitation using a second harmonic of a Ti:Sapphire femtosecond oscillator with pulse duration < 200 fs, which operates at a fundamental wavelength of ~900 nm. The average excitation power was kept at ~26 μW. The experimental data was fitted using a bi-exponential decay function:

$$y = y_0 + A_1 exp\left(\frac{-(x - x_0)}{t_1}\right) + A_2 exp\left(\frac{-(x - x_0)}{t_2}\right)$$

where, $y_0$ = offset, $x_0$ = center, $A_1$ & $A_2$ = amplitude, and $t_1$ & $t_2$ = time constant ($t_2 > t_1$). Fitted decay time constants are as follows,

$MoS_2$ in 1L HS: $t_1 = \sim 10.25 \pm 0.71$ ps, $t_2 = \sim 28.84 \pm 3.05$ ps

$WS_2$ in 1L HS : $t_1 = \sim 11.01 \pm 0.36$ ps, $t_2 = \sim 33.24 \pm 1.21$ ps

$WS_2$ in 2L HS : $t_1 = \sim 10.73 \pm 0.32$ ps, $t_2 = \sim 42.40 \pm 0.91$ ps

**Supporting information:**

The supporting information is available online: atomic force microscopy scan, PL emission and analysis, TR-PL measurement and details of the theory calculations.

**Declarations:**

**Data and code availability**

All the necessary data are presented in the manuscript and supplementary file. Technical details of the theoretical calculations are available from the corresponding author upon reasonable request.

**Acknowledgments**

This work is supported by the National Science Centre, Poland (Grant No. 2022/47/D/ST3/02086). K.W. and T.T. acknowledge support from the JSPS KAKENHI (Grant No. 19H05790, 20H00354, and 21H05233). D.D. and S.K.N. would like to appreciate the funding by the DST project (RP-312), and also acknowledge the access of the high-performance computing facility provided by the Institute of Physics, Bhubaneswar and Indian Institute of Technology Bhubaneswar. We thank Karol Nogajewski at the Center of New Technologies (CeNT) in the University of Warsaw for helping in using the e-beam evaporator.

**Author contributions:**

A.K. conceived the project, designed the experiments and provided the necessary funding. A.K. and G. fabricated the samples. G., T.K. and N.Z. performed the optical measurements. G. and A.K. carried out the data analysis. M.A., D.D., S.A. and S.K.N. conducted the theoretical studies. A.K., G., A.B. and M.R.M. decoded the experimental results. T.T. and K.W provided the hBN crystals. A.K., M.A. and G. wrote the manuscript. All the authors were consulted and feedback were taken before the submission.

**Competing interests**

We declare no competing financial interests.

**Note**

We did not use any A.I. software in auto-generating texts.

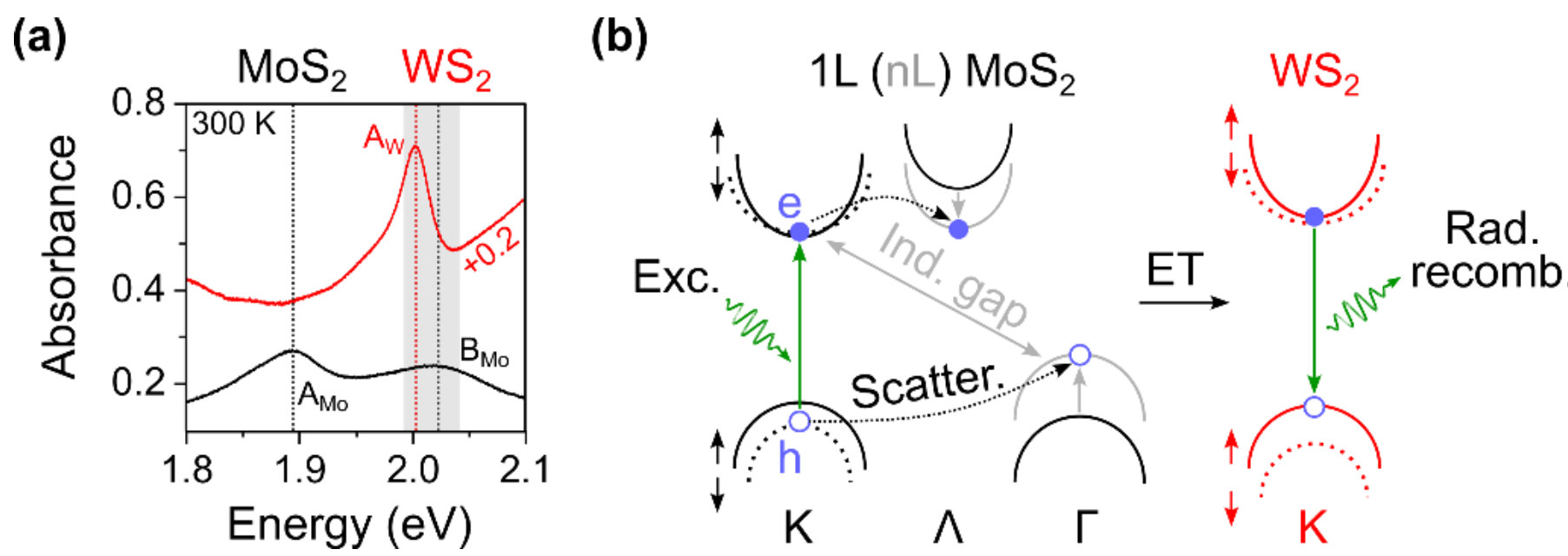

Figure 1: (a) Room temperature (RT) absorption spectra of 1Ls $MoS_2$ and $WS_2$, calculated from the RC experiment. Shaded area shows an almost 'resonant' overlap between the $MoS_2$ B ($B_{Mo}$) and $WS_2$ A ($A_W$) excitonic peaks. A constant background was added to $WS_2$ spectrum for a better visual representation. (b) Electronic spin-resolved sketch of the different inter-/intralayer processes: resonant excitation at $MoS_2$ B excitonic level create *e-h* pairs in the $MoS_2$ K valley. These *e-h* pairs can scatter to the Λ/Γ valleys, respectively depending on the layer dependent direct-to-indirect bandgap transition and further to the $WS_2$ A excitonic level *via* interlayer energy transfer (ET) process. The intensity of the $WS_2$ emission is determined by the efficiency of the different inter-/intravalley transitions in the $MoS_2$ layer.

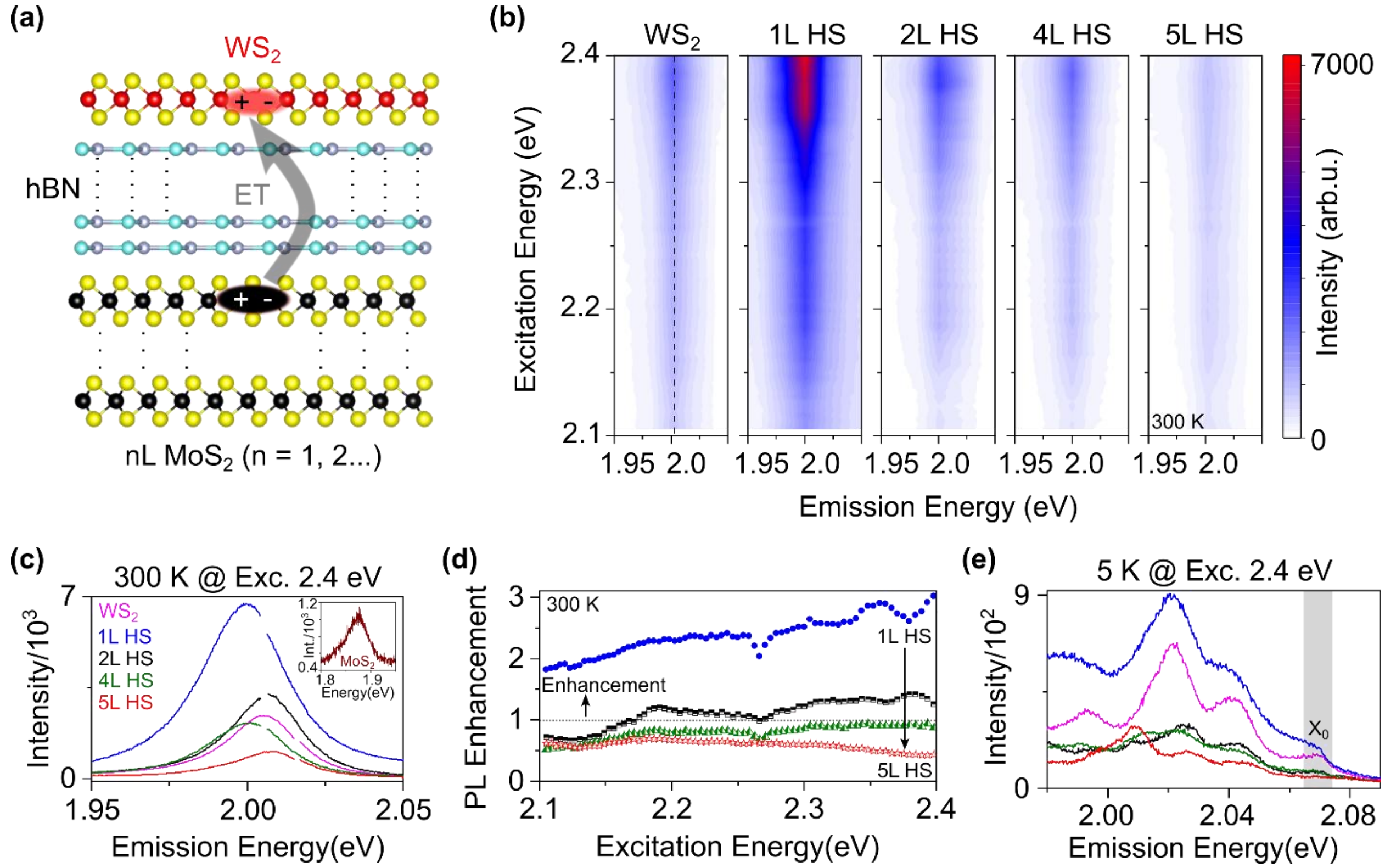

Figure 2: (a) Graphical representation of the 1L $WS_2$ and nL $MoS_2$ (n = 1, 2, 4 & 5) HS separated by thin hBN layer. This structure was placed onto a thick hBN/$SiO_2$(90 nm)/Si substrate. We tune the $MoS_2$ bandgap by increasing the layer number to observe the effect of the 'reverse' ET from the $MoS_2$ B excitonic level to the $WS_2$ A exciton. (b) Photoluminescence excitation (PLE) false colormaps from the $WS_2$ and HSs areas taken at 300 K. HS areas are labelled with 1L, 2L etc. to distinguish between the different $MoS_2$ layer numbers. All the maps are plotted with the same intensity scale for a better visual comparison. (c) RT photoluminescence (PL) spectra from all the regions under 2.4 eV laser excitation. Inset shows the $MoS_2$ PL emission. 1L HS area shows ~3× brighter PL emission as compared to the isolated $WS_2$ region. (d) PL enhancement factor as a function of the excitation energy calculated from the PLE maps shown in (b). To calculate the enhancement factor, PL intensities from the HS areas are divided by the $WS_2$ emission (along the vertical dashed line in (b)). 1L HS area shows an increasing PL emission with the excitation energy and the 2L HS area shows a similar behavior after ~2.17 eV excitation. While, the 4L and 5L HS areas show PL quenching throughout the entire excitation energy. (e) Low temperature (LT) (5 K) PL spectra from all the regions under 2.4 eV excitation energy. $WS_2$ neutral exciton ($X_0$) shows a slight enhancement only in the 1L HS area (marked by the shaded region). To differentiate the emissions from the individual regions a similar color scheme is used as in (c).

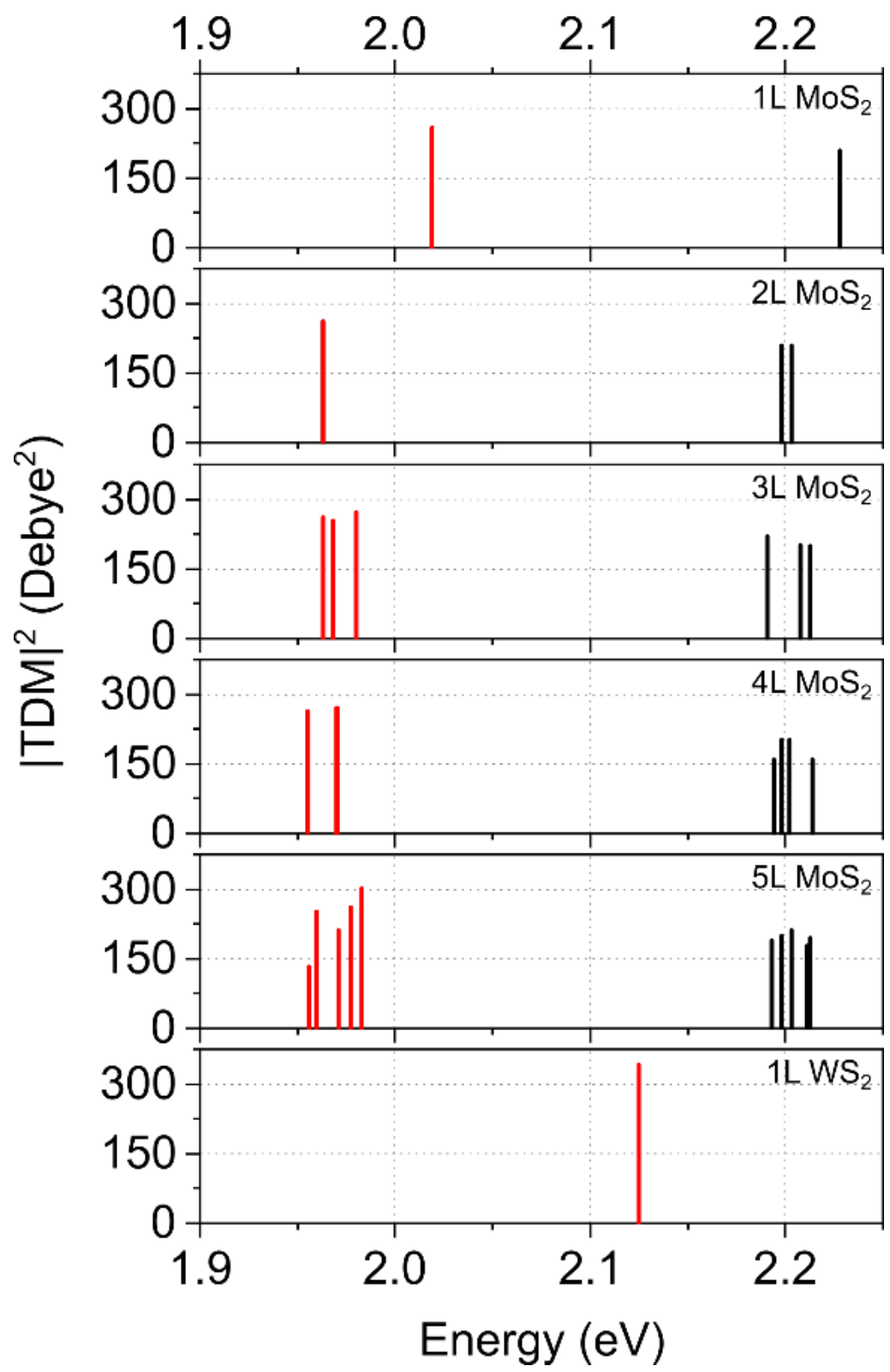

Figure 3: The squared transition dipole moments (TDM) at K valley is plotted along with the energy difference between the corresponding bands of 1L to 5L $MoS_2$ and 1L $WS_2$, respectively. Red and black bars correspondingly represent the A and B excitonic transitions.

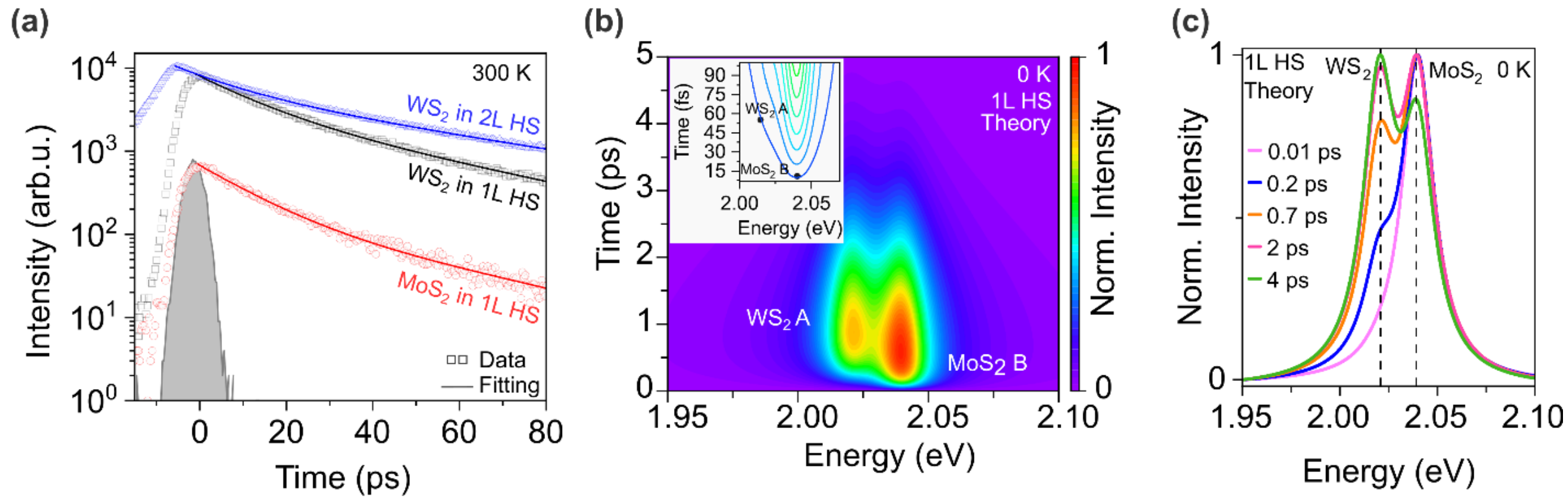

Figure 4: (a) RT time resolved PL (TR-PL) spectra of the $MoS_2$ and $WS_2$ from the 1L HS, and $WS_2$ in the 2L HS region measured at 300K. The instrument response function is shown by the shaded curve. Experimental decay profiles were fitted using a bi-exponential decay function. Fitted decay time for $MoS_2$, $WS_2$ in 1L HS and $WS_2$ in 2L HS are ~10.25 ± 0.71 ps, 28.84 ± 3.05 ps; 11.01 ± 0.36 ps, 33.24 ± 1.21 ps; and 10.73 ± 0.32 ps, 42.40 ± 0.91 ps, respectively. (b) Calculated false colormap of the time- and energy-dependent PL emission from the 1L HS at 0 K. Inset is the zoomed in snapshots to show the evolution of the $MoS_2$ B and $WS_2$ A states' emissions at early timescale. (c) Time evolution of the PL emission from the 1L HS (horizontal cuts from (b)). Initially $MoS_2$ B emission dominates in the HS. Then over the time $WS_2$ A emission starts to dominate.